\documentclass[twocolumn,showpacs,preprintnumbers,amsmath,amssymb,prb]{revtex4}
\usepackage[pdftex]{graphicx}% Include figure files
\usepackage{dcolumn}% Align table columns on decimal point
\usepackage{bm}% bold math

\begin{document}

\title{Temporal Monitoring of Non-resonant Feeding of Semiconductor Nanocavity Modes by Quantum Dot Multiexciton Transitions}% Force line breaks with \\

\author{A. Laucht}
\thanks{These authors contributed equally to this work.}
\author{M. Kaniber}
\thanks{These authors contributed equally to this work.}
\author{A. Mohtashami}
\author{N. Hauke}
\author{M. Bichler}
\author{J. J. Finley}%
\email{finley@wsi.tum.de}
\affiliation{Walter Schottky Institut, Technische Universit\"at M\"unchen, Am Coulombwall 3, 85748 Garching, Germany}%

\date{\today}% It is always \today, today,

\begin{abstract}
We experimentally investigate the non-resonant feeding of photons into the optical mode of a zero dimensional nanocavity by quantum dot multiexciton transitions. Power dependent photoluminescence measurements reveal a super-linear power dependence of the mode emission, indicating that the emission stems from multiexcitons. By monitoring the temporal evolution of the photoluminescence spectrum, we observe a clear anticorrelation of the mode and single exciton emission; the mode emission quenches as the population in the system reduces towards the single exciton level whilst the intensity of the mode emission tracks the multi-exciton transitions. Our results lend strong support to a recently proposed mechanism \cite{Winger09} mediating the strongly non-resonant feeding of photons into the cavity mode.
\end{abstract}

%Use showkeys class option if keyword
\maketitle
%INTRODUCTION
Semiconductor quantum dots (QD) exhibit atom like properties such as a discrete interband optical spectrum with nearly homogeneously broadened transitions and shell filling effects as carriers (electrons and holes) are added. As a result, they are natural candidates for conducting cavity quantum electrodynamics (cQED) experiments in the solid-state~\cite{Vahala03}. Amongst the many potential applications of single QD-cavity structures are the efficient and deterministic generation of indistinguishable photons~\cite{Santori02}, devices that exploit single photon quantum non-linearities~\cite{Englund07}, and ultra low threshold nanolasers~\cite{Strauf06}. Whilst exhibiting many properties known from atom based cQED, a number of surprising deviations from this model system have been identified. For example, recent experiments have revealed pronounced emission from the cavity mode, even when all the discrete QD emission lines are spectrally detuned~\cite{Hennessy07, Press07, Kaniber08b}. This effect has been attributed to phonon mediated dot-cavity interactions for small dot-cavity detunings up to a few meV~\cite{wilson-rae02, suffcynski09, Hohenester09, ota09}. However, non-resonant QD-cavity coupling was also observed for much larger detunings, up to $\sim20$~meV~\cite{Hennessy07, Kaniber08b, Winger09} which cannot be attributed to phonon mediated processes. 
Experimental investigations of such highly non-resonant coupling revealed apparently contradictory observations in photon auto and cross correlation measurements~\cite{Winger09, Hennessy07, Press07, Kaniber08b}:
%Recently, Winger \textit{et al.}~\cite{Winger09} presented experimental investigations of such highly non-resonant coupling, whereupon photon auto and cross correlation measurements revealed several, apparently contradictory observations: 
(i) the QD emission is strongly anti-bunched, (ii) the cavity emission is Poissonian whilst (iii) pronounced cross correlations exist between the QD and cavity mode emission. In Winger \textit{et al.}~\cite{Winger09}, these observations were explained by a model that links the strong off-resonant cavity mode emission to Purcell enhanced decay into a quasicontinuum of few particle states. The coexistence of the discrete QD states, where all particles are confined in the dot, and the quasicontinuum, where some particles occupy states in the wetting layer is inherent to the mesoscopic nature of the QD confinement.

%In this letter
In this article, we present experimental investigations of cavity mode feeding from spectrally detuned QD multiexciton states for a wide range of negative and positive single exciton - mode detunings ($\lambda_{QD}-\lambda_{cav}=-9$~nm to $+14$~nm). Both continuous wave (cw) and time resolved spectroscopy measurements are performed to probe the origin of the non-resonant cavity mode emission. Our results strongly support the identification of the dot-cavity coupling mechanism presented in Ref. ~\cite{Winger09} and, furthermore, allow us to directly monitor the temporal evolution of the emission spectrum as the level of excitation in the system reduces. Our measurements clearly establish a direct link between the mode emission and multiexciton states of the QD and are, therefore, in excellent agreement with the results presented in Ref.~\cite{Winger09}.

\begin{figure}[t!]
\includegraphics[width=0.9\columnwidth]{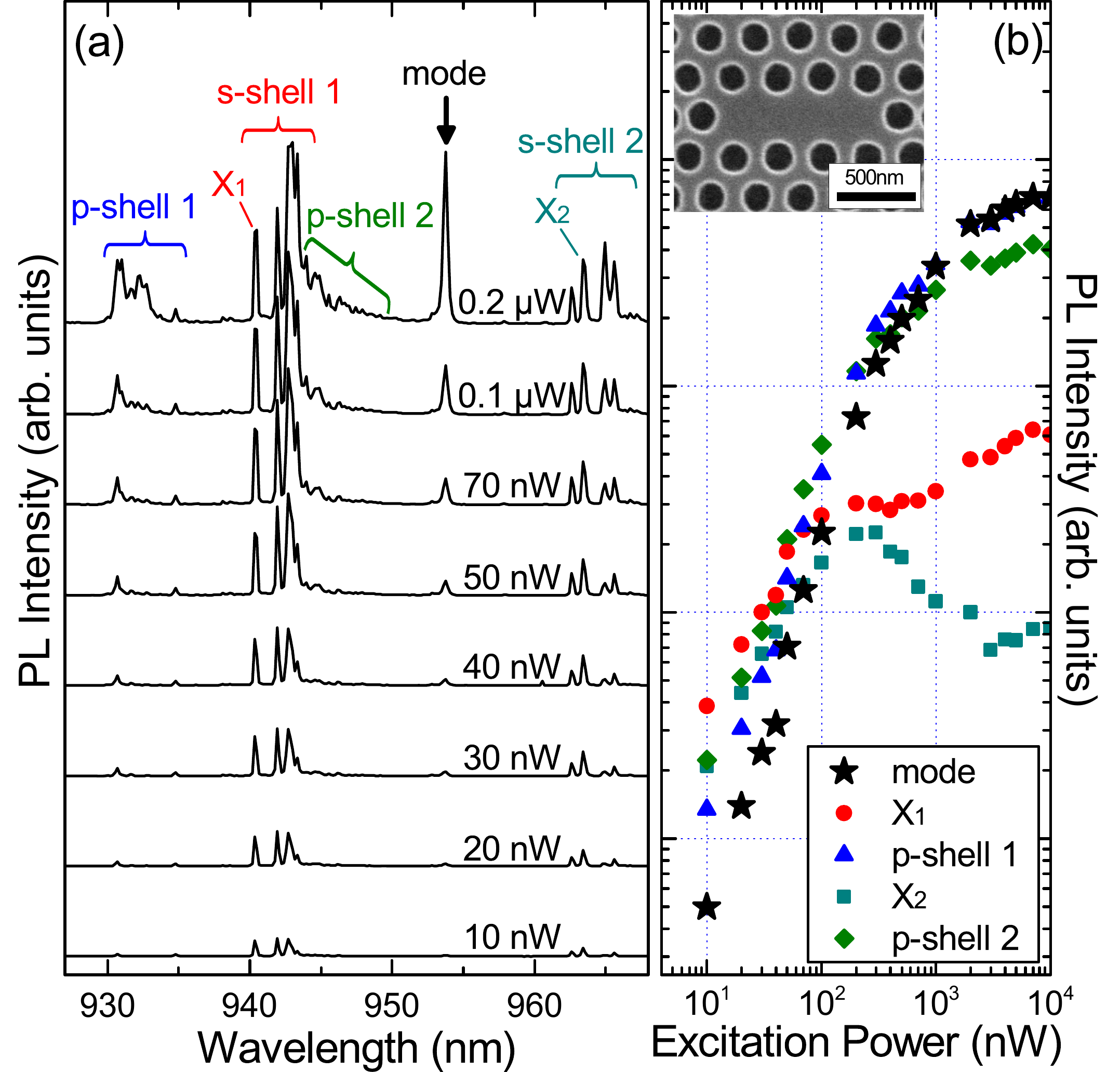} %0.9
\caption{\label{figure1} (color online) (a) Power dependent PL spectra recorded from a QD cavity system with two QDs, excitation level increasing from bottom to top. (b) Integrated PL intensity for the cavity mode (black stars), selected single exciton transitions from the s-shell of QD1 (labelled X$_{1}$ - red circles) and from the s-shell of QD2 (labelled X$_{2}$ - cyan rectangles). At higher power additional emission is observed from the p-shell multiexcitons of QD1 (blue triangles) and QD2 (green diamonds). (inset) Scanning electron microscopy image of the photonic crystal L3 nanocavity.}
\end{figure}

% SAMPLE FABRICATION
%
The sample investigated was grown by molecular beam epitaxy and consisted of the following layers grown on an semi-insulating GaAs substrate: A $500$~nm thick Al$_{0.8}$Ga$_{0.2}$As sacrificial layer, followed by a $180$~nm thick GaAs slab waveguide. A single layer of In$_{0.5}$Ga$_{0.5}$As QDs was incorporated at the midpoint of this waveguide. A two-dimensional photonic crystal formed by a triangular array of air holes was realized using a combination of electron-beam lithography and reactive ion etching. Nanocavities were established by introducing point defects consisting of three missing holes in a row (L3 cavity)~\cite{Akahane03} as shown in the scanning electron microscopy image in the inset of Fig.~\ref{figure1}~(b). In a final step, free standing GaAs membranes were formed by HF wet etching.

%
% EXPERIMENTAL SETUP
%
For optical characterization the sample was mounted in a liquid He-flow cryostat and cooled to $10-15$~K. For the excitation we either used a pulsed Ti-Sapphire laser ($80$~MHz repetition frequency, $2$~ps pulse duration) or a cw Ti-Sapphire laser tuned to an excitation wavelength of $\lambda_{exc}=850$~nm, spectrally in resonance with the wetting layer absorption continuum. The QD photoluminescence (PL) was collected via a $100\times$ microscope objective (NA=0.8) providing a spatial resolution of $\sim700$~nm. The signal was spectrally analyzed using a $0.5$~m imaging monochromator and detected using a Si-based, liquid nitrogen cooled CCD. For time-resolved spectroscopy we used a silicon avalanche photodiode with a temporal resolution of $\sim150$~ps after deconvolution with the system instrument response function (IRF).
%

%TEXT FOR FIGURE 1
%FIGURE 1
In Fig.~\ref{figure1}~(a) we show PL spectra of the investigated QD-cavity system as a function of the cw equivalent power of the excitation pulses used in our experiments. The power is increased from $10$~nW at the bottom spectrum to $200$~nW at the top spectrum. At low excitation levels we observe emission from the s-shells of two QDs (labelled QD1 and QD2) at $\lambda_{s-shell 1}=940-945$~nm and $\lambda_{s-shell 2}=962-967$~nm. Measurements with mode-resonant excitation (not shown here) demonstrated that QD1 is spatially well coupled to the electric field maximum of the investigated cavity mode, whilst QD2 is less strongly coupled~\cite{Nomura06,Kaniber09}. At higher excitation levels, we observe the emergence of p-shell emission from QD1 and QD2 at $\lambda_{p-shell 1}=930-935$~nm and $\lambda_{p-shell 2}=943-948$~nm, respectively. The fundamental mode of the L3 cavity is observed at $\lambda_{cav}=954$~nm, spectrally detuned from any of the discrete QD transitions. At low excitation powers ($<30$~nW) the cavity mode emission is hardly visible. In contrast, it dominates the spectrum at higher excitation powers, its onset occuring at an excitation level that is comparable to that needed to observe emission from the p-shell of QD1. This is already a first indication that feeding of the cavity mode is mediated by multiexciton QD transitions. 

For the transitions marked in Fig.~\ref{figure1}~(a), we extracted the integrated PL intensity and plot the power dependence in Fig.~\ref{figure1}~(b). The intensities of X$_1$ (red circles) and X$_2$ (cyan squares) exhibit a nearly linear dependence with exponents $m_{X1}=0.91\pm0.05$ and $m_{X2}=0.95\pm0.04$, respectively, proving the single exciton character of these lines~\cite{Finley01,abbarchi09}. The intensities of both transitions saturate at a power of $\sim100$~nW~\cite{Xsaturation,pshellint}. The p-shell emission intensity increases superlinearly with excitation power exhibiting exponents of $m_{p-shell 1}=1.53\pm0.07$ and $m_{p-shell 2}=1.42\pm0.07$, respectively. Furthermore, they saturate at a much higher laser power $>1$~$\mu$W, compared to the single exciton emission. In comparison with the s-shell and p-shell emission, the cavity mode intensity also increases superlinearly with an exponent of $m_{cav}=1.68\pm0.07$ (black stars). Furthermore, the mode emission saturates at an excitation power comparable to that observed for the QD p-shell transitions. These observations strongly suggest that the cavity mode emission is related to the multiexciton emission from the QDs.

%TEXT FOR FIGURE 2
%FIGURE 2
\begin{figure}[t!]
\includegraphics[width=0.9\columnwidth]{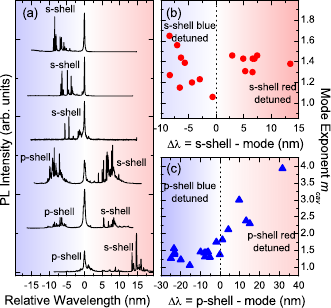} %0.9
\caption{\label{figure2} (color online) (a) Selected PL spectra of six different QD-cavity systems. (b),(c) The exponents of the extracted cavity mode emission intensity obtained from power dependent PL measurements of different QD-cavity systems plotted against the detuning relative to the s-shell (b) and p-shell (c).}
\end{figure}
%
%FIGURE 3
\begin{figure*}[t!]
\includegraphics[width=0.8\textwidth]{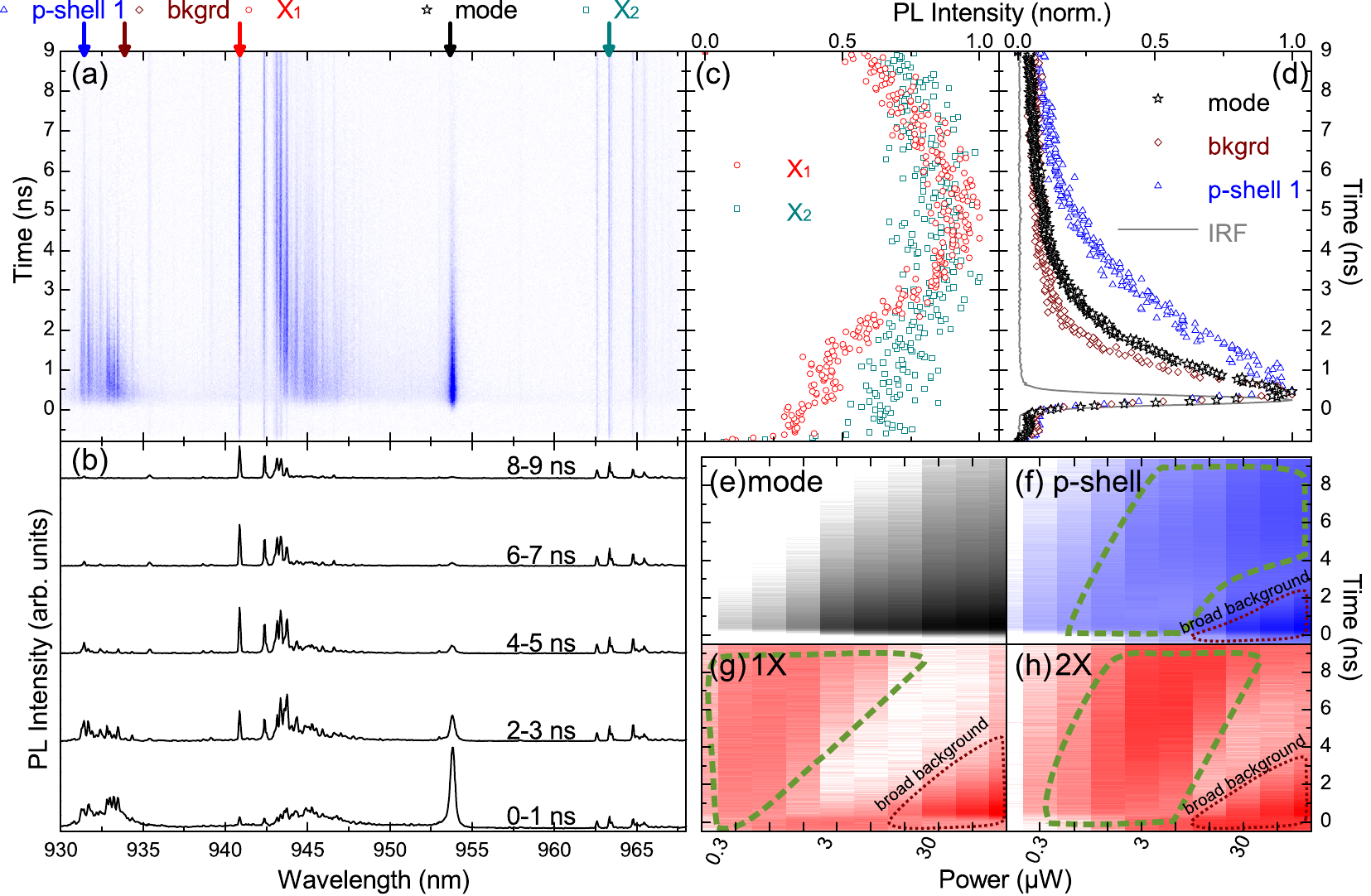} %1.6
\caption{\label{figure3} (color online) (a) False colour contour plot of the time-resolved PL intensity of a QD-cavity system as a function of emission wavelength~\protect\cite{remarkplot}. (b) PL spectra at different time delays after the laser pulse, each integrated over $1$~ns. (c),(d) Extracted, normalized PL intensity for (c) X$_1$ (red circles) and X$_2$ (cyan rectangles), and (d) cavity mode (black stars), p-shell 1 (blue triangles), and p-shell 1 background (dark red diamonds) as a function of time delay after the laser pulse. The IRF is plotted as gray solid line. (e-h) False colour contour plot of the time-resolved PL intensity (logarithmic scale) as a function of excitation power for (e) the cavity mode, (f) a p-shell state, (g) the single exciton, and (h) the biexciton of QD1.}
\end{figure*}

Consequently, we conducted similar power dependent measurements for various different QD-cavity systems and for different QD-cavity mode detunings. A selection of six representative spectra recorded from different QD-cavity systems is presented in Fig.~\ref{figure2}~(a). The data is presented on a relative wavelength scale with respect to the cavity mode emission. For the uppermost spectrum, the s-shell of the QD is on the short wavelength side of the mode (blue detuned), but for the bottom spectrum it is on the long wavelength side of the mode (red detuned). In the latter case, the mode is in resonance with the p-shell emission of the investigated QD. We extract the exponent of the power dependent mode PL intensity (c.f. Fig.~\ref{figure1}~(b)) and plot it as a function of detuning to the QD s-shell and p-shell in Fig.~\ref{figure2}~(b) and (c), respectively. When the cavity mode is detuned from the s-shell, the mode intensity exhibits a superlinear power dependence with exponents of $m_{cav}=1.2-1.6$, similar to our findings for the exponents of multiexciton transitions. Only when the mode is in resonance with the s-shell ($-5<\Delta\lambda<1$~nm in Fig.~\ref{figure2}~(b)), in a range where coupling via acoustic phonons dominates~\cite{Hohenester09}, do the exponents decrease to $m_{cav}=1.0-1.2$, characteristic of single exciton emission. In Fig.~\ref{figure2}~(c), we plot the measured exponents of the mode emission as a function of p-shell detuning. Providing that the mode is in resonance with the p-shell or at lower energies (p-shell blue detuned, $\Delta\lambda<2$~nm in Fig.~\ref{figure2}~(c)), we observe a superlinear behaviour with exponents between $m_{cav}=1.0$ and $m_{cav}=1.7$. For these detunings p-shell states have excess energy and can effectively feed the cavity mode. However, when the p-shell transitions are red detuned ($\Delta\lambda>2$~nm), the extracted exponents show values of $m_{cav}>2.0$ with the highest value being $m_{cav}=3.9$ measured at a detuning of $\Delta\lambda\sim+32$~nm. Here, feeding occurs from higher QD transitions and most likely from QD-wetting layer transitions. In summary, this behaviour strongly indicates multiexciton feeding of the mode with additional contribution of phonon assisted processes for small detunings, as reported in Ref. \cite{wilson-rae02, suffcynski09, Hohenester09, ota09}.

%TEXT FOR FIGURE 3

An unambiguous proof for the multiexciton feeding of the cavity mode is obtained from time-resolved PL measurements when the QD-cavity system is excited above saturation of the s-shell levels ($P=250$~nW). In Fig.~\ref{figure3}~(a) we plot the time-resolved emission intensity of the QD-cavity system introduced in Fig.~\ref{figure1} as a function of wavelength. Emission from the p-shell transitions of QD1 ($\lambda_{p-shell 1}=930-935$~nm) and QD2 ($\lambda_{p-shell 2}=943-948~$nm) occurs rapidly after the arrival of the laser excitation pulse and decays within a few nanoseconds~\cite{lifetime}. Emission from the s-shell transitions of QD1 ($\lambda_{s-shell 1}=940-945$~nm) and QD2 ($\lambda_{s-shell 2}=962-967$~nm) is temporally delayed, such that the maximum intensity is not reached until $\sim4-6$~ns after excitation. At that time the population in the dot has already decayed. Emission prior to the arrival of the laser pulse at $0$~ns originates from excitation due to the previous excitation cycle, $12.5$~ns earlier. The emission from the cavity mode at $\lambda_{cav}=954$~nm occurs rapidly after arrival of the excitation pulse and decays quickly within $\sim2$~ns. 

The spectra plotted in Fig.~\ref{figure3}~(b) show again the time evolution of the whole QD-cavity system. Here, we integrate the measured PL signal over $\Delta t=1$~ns time intervals and present the resulting spectra for the time intervals from $0-1$~ns, $2-3$~ns, $4-5$~ns, $6-7$~ns, and $8-9$~ns, from bottom to top in Fig.~\ref{figure3}~(b). For the first time interval, emission from the p-shell states of QD1 and QD2 and from the cavity mode dominates the spectrum. However, the intensity of this emission decreases rapidly and vanishes almost completly by $5$~ns after the excitation. The dominating peaks of the spectrum are now the s-shell emission of QD1 and QD2, whilst hardly any signal from the cavity mode is observed. 

For a more quantitative comparison, we plot the integrated, normalized PL intensity of X$_1$ (red circles) and X$_2$ (cyan rectangles) in Fig.~\ref{figure3}~(c), and of mode (black stars), p-shell 1 (blue triangles) and p-shell background emission (dark red diamonds) in Fig.~\ref{figure3}~(d). The solid gray line is the IRF of our experimental setup, measured with the detection tuned to the laser wavelength. It serves as reference for the time when the laser pulse excites the sample and allows us to determine the zero point of the time axis. While emission from the cavity mode, p-shell states and background occurs immediately after the laser pulse, the emission from the single excitons X$_1$ and X$_2$ is delayed and is temporally completely uncorrelated with the mode emission. It is also interesting to take a closer look at the decay times. The cavity mode decays ($\tau_{mode}=1.4\pm0.1$~ns) even faster than the selected discrete p-shell state ($\tau_{p-shell1}=2.2\pm0.1$~ns), but slower than the p-shell background between the discrete emission lines ($\tau_{bkgrd}=0.8\pm0.1$~ns). Moreover, feeding of the cavity mode apparently occurs from many different multiexciton states as well as from the broad background which is present at higher excitation levels~\cite{Dekel98}.

Finally, in Fig.~\ref{figure3}~(e-h) we present time-resolved measurements of the same QD-cavity system at different levels of the cw equivalent excitation power. The different panels show the decay transients measured at the wavelength of the mode (e), of a transition in the p-shell of QD1 (f), a single exciton transition (g), and the biexciton (h) of QD1 as a function of the excitation power. Here, the white color corresponds zo zero emission intensity. 
The exciton is the longest-living state and its emission shifts rapidly to later times as the excitation power is increased (the dashed, green line acts as a guide to the eye), since the population in the dot has to decay first. The prompt emission observed at short times for high powers (dotted, dark red line) originates from the broadband emission that is always present for very strong pumping. For the biexciton the delay due to state filling effects is not as strongly pronounced as for the exciton, and for the p-shell state it is again smaller, as expected. However, all three time resolved - excitation power plots exhibit a comparable contribution from the broad background at the highest powers investigated. Again, we can relate these emission characteristics to the emission from the cavity mode. We observe that the mode emits faster than all the other states and no delay at high powers can be recognized. There is no visible correlation with the single exciton and the biexciton state for this detuning. Most of the emission appears even faster than that of the p-shell and shows a good temporal accordance to the background emission.

%%%CONCLUSION
In conclusion, we have presented experimental studies of the non-resonant feeding of semiconductor nanocavity modes by multiexciton states. The power dependence of the mode reflects the superlinear behaviour of these states, while time-resolved spectroscopy experiments above saturation of the s-shell states show a temporal correlation with the multiexcitonic background. Our results strongly support a model of the non-resonant emission as arising from Purcell enhanced decay into a multi-exciton final state continuum~\cite{Winger09}.

%%%Acknowledgements
We would like to thank U. Hohenester and A. Imamo\u{g}lu for fruitful discussions. We acknowledge financial support of the DFG via the SFB 631, Teilprojekt B3 and the German Excellence Initiative via NIM.

\bibliography{Papers,footnotes}% Produces the bibliography via BibTeX.

\begin{thebibliography}{24}
\expandafter\ifx\csname natexlab\endcsname\relax\def\natexlab#1{#1}\fi
\expandafter\ifx\csname bibnamefont\endcsname\relax
  \def\bibnamefont#1{#1}\fi
\expandafter\ifx\csname bibfnamefont\endcsname\relax
  \def\bibfnamefont#1{#1}\fi
\expandafter\ifx\csname citenamefont\endcsname\relax
  \def\citenamefont#1{#1}\fi
\expandafter\ifx\csname url\endcsname\relax
  \def\url#1{\texttt{#1}}\fi
\expandafter\ifx\csname urlprefix\endcsname\relax\def\urlprefix{URL }\fi
\providecommand{\bibinfo}[2]{#2}
\providecommand{\eprint}[2][]{\url{#2}}

\bibitem[{\citenamefont{Winger et~al.}(2009)\citenamefont{Winger, Volz, Tarel,
  Portolan, Badolato, Hennessy, Hu, Beveratos, Finley, Savona
  et~al.}}]{Winger09}
\bibinfo{author}{\bibfnamefont{M.}~\bibnamefont{Winger}}, et~al.,
  \bibinfo{journal}{Phys. Rev. Lett.} \textbf{\bibinfo{volume}{103}},
  \bibinfo{pages}{207403} (\bibinfo{year}{2009}).

\bibitem[{\citenamefont{Vahala}(2003)}]{Vahala03}
\bibinfo{author}{\bibfnamefont{K.~J.} \bibnamefont{Vahala}},
  \bibinfo{journal}{Nature} \textbf{\bibinfo{volume}{424}},
  \bibinfo{pages}{839} (\bibinfo{year}{2003}).

\bibitem[{\citenamefont{Santori et~al.}(2002)\citenamefont{Santori, Fattal,
  Vuckovi\'{c}, Solomon, and Yamamoto}}]{Santori02}
\bibinfo{author}{\bibfnamefont{C.}~\bibnamefont{Santori}}, et~al.,
  \bibinfo{journal}{Nature} \textbf{\bibinfo{volume}{419}},
  \bibinfo{pages}{594} (\bibinfo{year}{2002}).

\bibitem[{\citenamefont{Englund et~al.}(2007)\citenamefont{Englund, Faraon,
  Fushman, Stoltz, Petroff, and Vuckovic}}]{Englund07}
\bibinfo{author}{\bibfnamefont{D.}~\bibnamefont{Englund}}, et~al.,
  \bibinfo{journal}{Nature} \textbf{\bibinfo{volume}{450}},
  \bibinfo{pages}{857} (\bibinfo{year}{2007}).

\bibitem[{\citenamefont{Strauf et~al.}(2006)\citenamefont{Strauf, Hennessy,
  Rakher, Choi, Badolato, Andreani, Hu, Petroff, and Bouwmeester}}]{Strauf06}
\bibinfo{author}{\bibfnamefont{S.}~\bibnamefont{Strauf}}, et~al.,
  \bibinfo{journal}{Phys. Rev. Lett.} \textbf{\bibinfo{volume}{96}},
  \bibinfo{pages}{127404} (\bibinfo{year}{2006}).

\bibitem[{\citenamefont{Hennessy et~al.}(2007)\citenamefont{Hennessy, Badolato,
  Winger, Gerace, Atature, Gulde, Falt, Hu, and Imamo\u{g}lu}}]{Hennessy07}
\bibinfo{author}{\bibfnamefont{K.}~\bibnamefont{Hennessy}}, et~al.,
  \bibinfo{journal}{Nature} \textbf{\bibinfo{volume}{445}},
  \bibinfo{pages}{896} (\bibinfo{year}{2007}).

\bibitem[{\citenamefont{Press et~al.}(2007)\citenamefont{Press, Gotzinger,
  Reitzenstein, Hofmann, Loffler, Kamp, Forchel, and Yamamoto}}]{Press07}
\bibinfo{author}{\bibfnamefont{D.}~\bibnamefont{Press}}, et~al.,
  \bibinfo{journal}{Phys. Rev. Lett.} \textbf{\bibinfo{volume}{98}},
  \bibinfo{pages}{117402} (\bibinfo{year}{2007}).

\bibitem[{\citenamefont{Kaniber
  et~al.}(2008{\natexlab{a}})\citenamefont{Kaniber, Laucht, Neumann,
  Villas-Boas, Bichler, Amann, and Finley}}]{Kaniber08b}
\bibinfo{author}{\bibfnamefont{M.}~\bibnamefont{Kaniber}}, et~al.,
  \bibinfo{journal}{Phys. Rev. B} \textbf{\bibinfo{volume}{77}},
  \bibinfo{pages}{161303(R)} (\bibinfo{year}{2008}{\natexlab{a}}).

\bibitem[{\citenamefont{Wilson-Rae and Imamoglu}(2002)}]{wilson-rae02}
\bibinfo{author}{\bibfnamefont{I.}~\bibnamefont{Wilson-Rae}} et~al.,
  \bibinfo{journal}{Phys. Rev. B} \textbf{\bibinfo{volume}{65}},
  \bibinfo{pages}{235311} (\bibinfo{year}{2002}).

\bibitem[{\citenamefont{Suffcynski et~al.}(2009)\citenamefont{Suffcynski,
  Dousse, Lemaitre, Sagnes, Lanco, Bloch, Voisin, and Snellart}}]{suffcynski09}
\bibinfo{author}{\bibfnamefont{J.}~\bibnamefont{Suffcynski}}, et~al.,
  \bibinfo{journal}{Phys. Rev. Lett.} \textbf{\bibinfo{volume}{103}},
  \bibinfo{pages}{027401} (\bibinfo{year}{2009}).

\bibitem[{\citenamefont{Hohenester et~al.}(2009)\citenamefont{Hohenester,
  Laucht, Kaniber, Hauke, Neumann, Mohtashami, Seliger, Bichler, and
  Finley}}]{Hohenester09}
\bibinfo{author}{\bibfnamefont{U.}~\bibnamefont{Hohenester}}, et~al.,
  \bibinfo{journal}{Phys. Rev. B} \textbf{\bibinfo{volume}{80}},
  \bibinfo{pages}{201311(R)} (\bibinfo{year}{2009}).

\bibitem[{\citenamefont{Ota et~al.}(2009)\citenamefont{Ota, Iwamoto, Kumagai,
  and Arakawa}}]{ota09}
\bibinfo{author}{\bibfnamefont{Y.}~\bibnamefont{Ota}}, et~al.,
  \bibinfo{journal}{arXiv.org:0908.0788}  (\bibinfo{year}{2009}).

\bibitem[{\citenamefont{Akahane et~al.}(2003)\citenamefont{Akahane, Asano,
  Song, and Noda}}]{Akahane03}
\bibinfo{author}{\bibfnamefont{Y.}~\bibnamefont{Akahane}}, et~al.,
  \bibinfo{journal}{Nature} \textbf{\bibinfo{volume}{425}},
  \bibinfo{pages}{944} (\bibinfo{year}{2003}).

\bibitem[{\citenamefont{Nomura et~al.}(2006)\citenamefont{Nomura, Iwamoto,
  Nakaoka, Ishida, and Arakawa}}]{Nomura06}
\bibinfo{author}{\bibfnamefont{M.}~\bibnamefont{Nomura}}, et~al.,
  \bibinfo{journal}{Jpn. J. Appl. Phys.} \textbf{\bibinfo{volume}{45}},
  \bibinfo{pages}{6091} (\bibinfo{year}{2006}).

\bibitem[{\citenamefont{Kaniber et~al.}(2009)\citenamefont{Kaniber, Neumann,
  Laucht, Bichler, Amann, and Finley}}]{Kaniber09}
\bibinfo{author}{\bibfnamefont{M.}~\bibnamefont{Kaniber}}, et~al.,
  \bibinfo{journal}{New J. Phys.} \textbf{\bibinfo{volume}{11}},
  \bibinfo{pages}{013031} (\bibinfo{year}{2009}).

\bibitem[{\citenamefont{Finley et~al.}(2001)\citenamefont{Finley, Ashmore,
  Lemaitre, Mowbray, Skolnick, Itskevich, Maksym, Hopkinson, and
  Krauss}}]{Finley01}
\bibinfo{author}{\bibfnamefont{J.~J.} \bibnamefont{Finley}}, et~al.,
  \bibinfo{journal}{Phys. Rev. B} \textbf{\bibinfo{volume}{6307}},
  \bibinfo{pages}{073307} (\bibinfo{year}{2001}).

\bibitem[{\citenamefont{Abbarchi et~al.}(2009)\citenamefont{Abbarchi,
  Mastrandrea, Kuroda, Mano, Vinattieri, Sakoda, and Gurioli}}]{abbarchi09}
\bibinfo{author}{\bibfnamefont{M.}~\bibnamefont{Abbarchi}}, et~al.,
  \bibinfo{journal}{J. Appl. Phys.} \textbf{\bibinfo{volume}{106}},
  \bibinfo{pages}{053504} (\bibinfo{year}{2009}).

\bibitem[{Xsa()}]{Xsaturation}
\bibinfo{note}{The increase of the intensity of X$_1$ for higher excitation
  powers is explained by the appearance of a background under very strong
  pumping. The unexpected reduction of the intensity of X$_2$, however, is
  attributed to the long lifetime ($\tau\sim13$~ns) emitting deep inside the
  photonic bandgap, comparable to the laser repetition period.}

\bibitem[{psh()}]{pshellint}
\bibinfo{note}{For the PL intensity of the p-shell transitions of QD1 and QD2
  we integrated over the p-shell emission band since individual transitions
  could not be well resolved.}

\bibitem[{rem()}]{remarkplot}
\bibinfo{note}{The spectrally broad mode emission at early times is an artifact
  of plotting and not a physical effect. Here, the linear colour scale was
  chosen such that features of low intensity are well visible and the intensity
  above a certain threshold is plotted with the same colour.}

\bibitem[{lif()}]{lifetime}
\bibinfo{note}{Here, the lifetime of all transitions is lengthened due to the
  two dimensional photonic bandgap~\cite{Kaniber07,Kaniber08a}.}

\bibitem[{\citenamefont{Dekel et~al.}(1998)\citenamefont{Dekel, Gershoni,
  Ehrenfreund, Spektor, Garcia, and Petroff}}]{Dekel98}
\bibinfo{author}{\bibfnamefont{E.}~\bibnamefont{Dekel}}, et~al.,
  \bibinfo{journal}{Phys. Rev. Lett.} \textbf{\bibinfo{volume}{80}},
  \bibinfo{pages}{4991} (\bibinfo{year}{1998}).

\bibitem[{\citenamefont{Kaniber et~al.}(2007)\citenamefont{Kaniber, Kress,
  Laucht, Bichler, Meyer, Amann, and Finley}}]{Kaniber07}
\bibinfo{author}{\bibfnamefont{M.}~\bibnamefont{Kaniber}}, et~al.,
  \bibinfo{journal}{Appl. Phys. Lett.} \textbf{\bibinfo{volume}{91}},
  \bibinfo{pages}{061106} (\bibinfo{year}{2007}).

\bibitem[{\citenamefont{Kaniber
  et~al.}(2008{\natexlab{b}})\citenamefont{Kaniber, Laucht, H\"{u}rlimann,
  Bichler, Meyer, Amann, and Finley}}]{Kaniber08a}
\bibinfo{author}{\bibfnamefont{M.}~\bibnamefont{Kaniber}}, et~al.,
  \bibinfo{journal}{Phys. Rev. B} \textbf{\bibinfo{volume}{77}},
  \bibinfo{pages}{073312} (\bibinfo{year}{2008}{\natexlab{b}}).

\end{thebibliography}

\end{document}